\documentclass[%
 reprint,
 amsmath,amssymb,
 aps,
]{revtex4-1}
\usepackage{physics}
\usepackage{lmodern}
\usepackage{xcolor}
\usepackage{url}
\usepackage{hyperref}
\usepackage{cleveref}
\usepackage{graphicx}
\usepackage{comment}

\usepackage[utf8]{inputenc}
\usepackage[T1]{fontenc}
\hypersetup{
  colorlinks=true,
  citecolor=magenta,
  linkcolor=blue,
  urlcolor=violet
 }
\DeclareUnicodeCharacter{2212}{\ensuremath{-}}

\def\be{\begin{equation}}\def\ee{\end{equation}}
\newcommand{\baa}{\begin{equation}\begin{aligned}}
\newcommand{\ea}{\end{aligned}\end{equation}}

\newcounter{para}
\newcommand\mypara{\par\refstepcounter{para}\textbf{\thepara .}\space}

\newdimen\hfuzz
\newdimen\vfuzz

\begin{document}

\title{\textbf{Coherent States in M-Theory: A Brane Scan using the Taub-NUT}}

\author{Joydeep Chakravarty$^1$}\email{joydeep.chakravarty@mail.mcgill.ca}
\author{Keshav Dasgupta$^1$}\email{keshav@hep.physics.mcgill.ca} 
\author{Diksha Jain$^2$}\email{diksha.jain@tifr.res.in}
\author{Dileep P. Jatkar$^{3,4}$}\email{dileep@hri.res.in}
\author{Archana Maji$^5$}\email{archana$\_$phy@iitb.ac.in}
\author{Radu Tatar$^6$}\email{Radu.Tatar@liverpool.ac.uk}

\affiliation{$^1$Department of Physics, McGill University, Montr\'eal, Qu\'ebec, H3A 2T8, Canada}

\affiliation{$^2$Theory Division, Tata Institute of Fundamental Research, Homi Bhabha Road, Mumbai 400 005, India}

\affiliation{$^3$Harish-Chandra Research Institute, Chhatnag Road, Jhunsi, Allahabad 211019,  India}

\affiliation{$^4$Homi Bhabha National Institute, Training School Complex, Anushaktinagar, Mumbai 400094, India}

\affiliation{$^5$Department of Physics, Indian Institute of Technology Bombay, Mumbai 400076, India}

\affiliation{$^6$Department of Mathematical Sciences, University of Liverpool, Liverpool, L69 7ZL, United Kingdom}

\begin{abstract}
The Taub-NUT geometry corresponds to the Kaluza-Klein monopole solution of M-theory and on dimension reduction along the Taub-NUT circle direction it becomes the D6 brane of type IIA string theory.  We show that the Taub-NUT geometry can be realised as a coherent state, or more appropriately as a Glauber-Sudarshan state in M-theory, once we take the underlying resurgence structure carefully.  Using the duality chain it in turn implies that all D-branes as well as NS5-branes can be realised as Glauber-Sudarshan states in string theory. Our analysis also leads to an intriguing possibility of realizing the gravity duals of certain non-conformal minimally-supersymmetric gauge theories by deforming a class of Glauber-Sudarshan states.
\end{abstract}

\maketitle

\newpage
\mypara{\textbf{Introduction:}} 
The non-perturbative physics of string theory captured by solitonic solutions is nicely characterised in terms of D-brane and NS5-brane solutions \cite{Dai, Polchinski:1995mt}.  Before the discovery of D-branes, solitonic p-brane solutions to supergravity equations of motion were obtained and attempts to study their dynamics using worldvolume field theory were also carried out.  One of the advantages of these supergravity approaches was that they were agnostic to the origin of supergravity theories.  In other words, the solutions were obtained not only in the type I or type II supergravities, which originate from underlying string theory but also in the eleven dimensional supergravity, which has its origin in the M-theory.

The stringy dualities have brought all string theories and M-theory on an equal footing, thereby bringing solitonic solutions in M-theory almost on par with the D(NS)-brane solutions.  While the latter enjoy the string worldsheet description, a microscopic description of solitons in M-theory continue to be an enigma.  Nonetheless, string dualities seamlessly relate M-theory solitons to string theory solitons \cite{witten1, Hull, Sennp}.  The soliton solution is typically a classical solution to non-linear equations of motion of the theory. 

It is believed that most of these soliton solutions can be expressed in terms of coherent states (see \cite{Dvali1} for a discussion on field theory solitons).  An explicit representation of this belief, however, is lacking in the literature.  It has been recently shown that de Sitter spacetime can be realised as a Glauber-Sudarshan(GS) state in type IIB string theory \cite{Brahma1, Brahma2, Brahma3}. The GS states differ from the usual coherent states $-$ which are constructed by shifting the {\it free} vacua $-$ by being constructed from shifting the {\it interacting} vacua. This construction begins with an M-theoretic configuration and using string dualities, one can obtain a solution in the type IIB theory.  A reader may get an impression that this is done by restricting to a sector corresponding to the consistent truncation, but that is not the case.  The GS state is constructed by taking into account terms at arbitrary orders in coupling as well as in the derivative expansion. (Keeping all the irrelevant operators exhaustively is possible because (a) we are at low energies where we know the degrees of freedom, and (b) an Exact Renormalization Group analysis \cite{Polchinski:1983gv} guarantees this.) The GS state describing the de Sitter geometry is a resurgent sum over contribution of all these terms \cite{Brahma4}.  

In this paper, we will demonstrate that the Taub-NUT geometry, which corresponds to the Kaluza-Klein(KK) monopole solution in M-theory, can be constructed as a GS coherent state over the Minkowski vacuum by doing a resurgent sum over all higher order terms. It is well known that the D6 brane of type IIA string theory in the strong coupling limit becomes the KK monopole solution in the M-theory.  On the other hand, an appropriate dimension reduction brings us to the type IIA KK monopole solution, which by duality chain gives us NS5-brane solution in both type IIA and IIB string theory.  Thus, using a combination of dimension reduction and stringy duality symmetries, our construction of the Taub-NUT space as a coherent state over the Minkowski vacuum of the M-theory provides a mechanism of showing that not only all D-branes but also the KK monopoles and the NS5 branes are GS coherent states in the appropriate string theory. Interestingly this also opens up an avenue to investigate gravity duals of certain non-conformal pure glue theories in the large ${\rm N}$ limit.
\mypara{\textbf{The Taub-NUT space is a Glauber-Sudarshan State:}} \label{TNisGS}
Let us quickly recall the KK monopole solution in M-theory.  In the low energy limit, physics of M-theory is completely captured by the 11D supergravity.  Massless fields of this theory are, the metric ${g}_{\rm MN}$, the three form gauge field ${\rm C}_{\rm MNP}$, and the gravitino fields $\Psi_{{\rm M}\alpha}$, and $\overline\Psi_{{\rm M}\alpha}$ where ${\rm M,N},\cdots = 0,\cdots\ ,9, 11$; and $\alpha=1,\cdots\ , 16$ are 11D vector and Dirac spinor indices respectively.  The KK monopole is a purely gravitational bosonic solution to the equations of motion, {\em i.e.}, the solution is completely characterised by the metric alone, with ${\rm C}_{\rm MNP}=0$, and $\Psi_{{\rm M}\alpha}= \overline\Psi_{{\rm M}\alpha} = 0$.  In other words, Taub-NUT is a solution to the vacuum Einstein equation of M-theory.
The KK monopole solution is described by the metric
\begin{align}
    ds^{2}&={g}_{\rm MN}dz^{\rm M} dz^{\rm N}=-dt^{2}+\sum_{m=1}^{6}dy^{m} dy^{m}+ds^{2}_{\text{TN}}\label{MthMetrIc}
\end{align}
where, $y^m$ denotes coordinates transverse to the Taub-NUT geometry $m=1,\cdots ,6$, and $ds^2_{\text{TN}}$ is the metric of the Euclidean Taub-NUT space \cite{townsend1995eleven}: 
\begin{align}
   ds^{2}_{\text{TN}}&= V(\mathbf{x})d\mathbf{x}\cdot d\mathbf{x}+V^{-1}(\mathbf{x})(dx^{11}+\mathbf{A}(\mathbf{x})\cdot d \mathbf{x})^2\ .
\end{align}
Here $x^{11}$ is the coordinate of the compact direction and $\mathbf{x}\equiv (x^7,x^8,x^9)$ are the three spatial coordinates transverse to the brane.  The functions $V(\mathbf{x})$ and $\mathbf{A}(\mathbf{x})$ are not independent, they are related to each other by the self-duality condition, 
\vspace{-2mm}
\begin{align}
\Vec{\nabla} \times \mathbf{A}=\vec{\nabla} V\implies \nabla^2 V=0 \ .
\label{eq:sdc} \end{align}
We will consider single centered solution which possesses spherical symmetry and corresponds to the following form of the function
\begin{equation}
    V(r) = 1 + \frac{\mu}{r}\ ,
\end{equation}
where $r=\sqrt{\mathbf{x}\cdot \mathbf{x}}$ and $\mu$ is a constant.  The singularity at $r=0$ is a coordinate singularity provided the coordinate $x^{11}$ has periodicity $4\pi \mu$.  The vector potential $\mathbf{A}$ can then be determined using the eq.\eqref{eq:sdc}.  It is worth emphasising that the KK monopole solution does not contain any harmonic function in the directions transverse to the Taub-NUT space.  While the Minkowski space is a solution with trivial holonomy, the Taub-NUT solution has $SU(2)$ holonomy and as a result in the supersymmetric set up, it preserves half the supersymmetry.  Thus the unit charge M-theory KK monopole solution in the polar coordinates can be written as
\begin{align}
    ds^{2}_{11}&=-dt^{2}+\sum_{m=1}^{6}dy^m dy^m+\left(1+\frac{\mu}{r}\right)(dr^2+r^2d\Omega_{2}^{2})\nonumber \\
    &+\left(1+\frac{\mu}{r}\right)^{-1}\left(dx^{11}+\mu \cos\theta d\varphi\right)^2\ ,\label{KKmoNopolEinMth}
\end{align}
The dimensional reduction of the compact coordinate $x^{11}$ leads to the ten dimensional metric, the dilaton and the Ramond-Ramond gauge field $\mathbf{A}$, which corresponds to the D6 brane solution in type IIA string theory.  The string coupling is given by
\begin{align}
    g_{s}=e^{\phi}&=\left(1+\frac{\mu}{r}\right)^{-3/4}\ ,
\end{align}
the radius of the compact eleventh direction is
\begin{align}\label{Mplgs}
    {\rm R}_{11}&=\sqrt{g_{11,11}}l_{p}=g_{s}^{2/3} l_{p}=g_{s}l_{s} \ ,
\end{align}
and $\mu=\frac{1}{2} g_sl_s = \frac{1}{2}g_{s}\sqrt{\alpha^\prime}$, implying its connection to the radius of the eleventh direction. 

We will now show that the Taub-NUT solution (\ref{KKmoNopolEinMth}) is a GS state in M-theory.  Our starting point will be the Minkowski solution to the M-theory equation of motion, namely,
\vspace{-2mm}
\begin{align}
    ds^{2}&=-dt^2+\sum_{m=1}^{6}dy^{m}dy^{m}+dr^2+ {r^{2}}d\widetilde\Omega_{2}^{2}\nonumber\\[-1mm]
    &+ {r^2} (dx^{11}+\mu \cos\theta d\varphi)^2\ ,\label{mink}
\end{align}
where we have chosen the Cartesian coordinates in six spatial directions; spherical polar coordinates in remaining four directions with the Hopf metric on $S^3$; and defined $d\widetilde\Omega_2^2 = \frac{d\theta^2}{4} + \mu^2 \sin^2\theta d\varphi^2$. The boundary condition will be satisfied because at $r\to {1\over k_{\rm IR}}$, where $k_{\rm IR}$ is the IR cutoff, the large radius eleventh circle will be replaced by a finite sized circle by the GS state as we shall see in 
\eqref{alpsphere} and \eqref{11circle}.

A general Euclidean path integral analysis in M-theory leads to the following expression for the expectation value of metric operator over the GS 
state \cite{Brahma4}:
\begin{widetext}
\begin{align}\label{resurmet}
\langle \delta {\bf g}_{\rm MN} \rangle_{\sigma} & \equiv \frac{\langle\sigma\vert \delta{\bf g}_{\rm MN}\vert \sigma\rangle}{\langle\sigma\vert \sigma\rangle} = 
\frac{\int [{\cal D} g_{\rm MN}] [{\cal D}{\rm C}_{\rm MNP}] [{\cal D}\overline\Psi_{\rm M}] [{\cal D}
\Psi_{\rm N}]~e^{-{\bf S}_{\rm tot}}~ \mathbb{D}^\dagger(\bar{\alpha}, \bar{\beta}, \bar\gamma) \delta g_{\rm MN}({\rm X}, t)
\mathbb{D}(\bar{\alpha}, \bar{\beta}, \bar\gamma) }{
\int [{\cal D} g_{\rm MN}] [{\cal D}{\rm C}_{\rm MNP}] [{\cal D}\overline\Psi_{\rm M}] [{\cal D}
\Psi_{\rm N}]
~e^{-{\bf S}_{\rm tot}} ~\mathbb{D}^\dagger(\bar{\alpha}, \bar{\beta}, \bar\gamma) 
\mathbb{D}(\bar{\alpha}, \bar{\beta}, \bar\gamma)}\nonumber\\
&=\Bigg[\frac{1}{ g^{1/l}}
\int_0^\infty d{\rm S} ~{\exp}\left(-\frac{\rm S}{ g^{1/l}}\right) \frac{1}{1 - {\cal A}{\rm S}^l}\Bigg]_{{\rm P. V}}\int_{k_{\rm IR}}^\mu d^{11}k ~\frac{\overline\alpha_{\rm MN}(k)}{a(k)}~
{\bf Re}\left(\psi_k({\rm X})~e^{-i k_0 t}\right)\ .
\end{align}
\end{widetext}
The subscript $\sigma$ of the metric expectation values is a mnemonic used to indicate that the expectation value is obtained in a specific Glauber-Sudarshan state $\vert\sigma\rangle = \mathbb{D}(\sigma)\vert\Omega\rangle \equiv \vert \bar\alpha, \bar\beta, \bar\gamma\rangle$. Here $\bar\alpha \equiv \bar\alpha_{\rm MN}, \bar\beta \equiv \bar\beta_{\rm MNP}, \bar\gamma \equiv \left(\Gamma_{{\rm M}\alpha},\bar\Gamma_{{\rm M}\alpha}\right)$ represent the GS states associated with the various components of the metric, fluxes and the Rarita-Schwinger fermions respectively in the Fourier space. $\mathbb{D}(\sigma)$ is the displacement operator acting on the {\it interacting} vacua, and therefore differs from the usual notion of the coherent state where the displacement operator acts on the {\it free} vacua. Additionally $\mathbb{D}^\dagger\mathbb{D} \ne \mathbb{D} \mathbb{D}^\dagger \ne 1$.  The bold-faced letters denote the operators, for example ${\bf g}_{\rm MN} \equiv {\bf g}^{(o)}_{\rm MN} + \delta {\bf g}_{\rm MN}$ is related to the metric operator such that 
$\frac{\langle\sigma\vert {\bf g}^{(o)}_{\rm MN}\vert \sigma\rangle}{\langle\sigma\vert\sigma\rangle}$ is the Minkowski background \eqref{KKmoNopolEinMth}. The other parameters appearing in \eqref{resurmet} are defined as follows. The coupling $g$ is related to the inverse power of ${\rm M}_p$ that appears as the suppression factor in a given derivative interaction (the fields are taken to be dimensionless), $l$ is the corresponding asymptotic Gevrey growth with ${\cal A}$ being related to the amplitudes of the so-called dominant {\it nodal diagrams}; and P. V is the principal value of the integral (see \cite{Brahma4} for details).  It is important to point out that this result incorporates resurgent sum over all higher order derivatives as well as curvature corrections. (In \cite{Brahma4} the analysis was done by mapping the various metric, flux and fermionic components to scalar degrees of freedom to avoid incorporating Faddeev-Popov ghosts, but the generic analysis, even including the derivative ghosts \cite{Brahma5}, while technically challenging may be done with some effort.)  One may even go beyond the dominant nodal-diagrams and incorporate NLO diagrams, including all possible higher powers of the derivative interactions in ${\bf S}_{\rm tot}$, but the result quoted in \eqref{resurmet} changes only by a constant multiplicative factor which may be absorbed in the definition of $\overline\alpha_{\rm MN}(k)$ \cite{Brahma5}. This surprising feature is of course a consequence of the wave-function renormalization that was predicted for a generic setting in \cite{Brahma1, Brahma2, Brahma3, Brahma4} and tells us that knowing $\overline\alpha_{\rm MN}(k)$ one would easily compute the expectation value $\langle {\bf g}_{\rm MN}\rangle_\sigma$ from the graviton propagator $a(k)$ and wave-function $\psi_k({\rm X})$, where ${\rm X}$ is the ten-dimensional spatial coordinate, in the energy range $k_{\rm IR} \le k \le \mu$ with $\Lambda > {\rm M}_p > {\rm M}_s > \mu > k_{\rm IR}$, where $k_{\rm IR}$ is the IR cut-off mentioned earlier. Alternatively, knowing the metric components, we can use the inverse transformation to compute $\overline\alpha_{\rm MN}(k)$, thus determining the GS state itself. Since the KK monopole solution is independent of the 3-form field ${\rm C}_{\rm MNP}$, we can consistently set it to zero, {\it i.e.} $\langle {\bf C}_{\rm MNP}\rangle_\sigma = 0$ (or $\bar\beta_{\rm MNP} = 0$), and concentrate only on the remaining bosonic field, namely the metric $\langle{\bf g}_{\rm MN}\rangle_\sigma$.  Furthermore, by implementing a change of variable $\mathrm{S}=u/\mathcal{A}^{1/l}$ we can simplify the principal value integral part of the expectation value of the metric in the following way:
\begin{align}
    \langle \delta {\bf g}_{\rm MN} \rangle_{\sigma} &=\frac{1}{c}\Bigg[
\int_0^\infty d{ u} ~{\rm exp}\left(-\frac{ u}{c}\right) \frac{1}{1 - { u}^l}\Bigg]_{{\rm P. V}} \nonumber \\
&\times\int_{k_{\rm IR}}^\mu d^{11}k ~\frac{\overline\alpha_{\rm MN}(k)}{a(k)}~
{\bf Re}\left(\psi_k({\rm X})~e^{-i k_0 t}\right)\ .\label{delmet}
\end{align}
where $c\equiv (\mathcal{A}g)^{1/l}$ and for convenience we will refer to the principal value integral, that is on the first line of eq.\eqref{delmet}, as $\textstyle\mathbb{I}_{c,l}$. (We will revisit this integral momentarily.) The expression in eq.\eqref{delmet} is for the fluctuation of a generic component of the metric and to obtain a specific metric, e.g., the Taub-NUT metric, we need to make a judicious choice of the Glauber-Sudarshan functions $\overline\alpha_{\rm MN}(k)$ such that the Fourier integral correctly reproduces desired metric components.

Our task is to find $\overline\alpha_{\rm MN}(k)$ such that we derive the Taub-NUT metric with following expectation values.
\begin{align}
    &\langle \mathbf{g}_{00}\rangle_{\sigma}=-1\ ,\ \ \langle \mathbf{g}_{11}\rangle_{\sigma}=1\ ,\ \ \langle \mathbf{g}_{22}\rangle_{\sigma}=1\ ,\ \cdots ,\nonumber \\
    &\langle \mathbf{g}_{66}\rangle_{\sigma}=1\ ,\ \ 
    \langle \mathbf{g}_{rr}\rangle_{\sigma}=V(r)\ ,\ \ \langle \mathbf{g}_{\theta \theta}\rangle_{\sigma}= r^{2}V(r)\ ,\nonumber \\
    &\langle \mathbf{g}_{\varphi\varphi}\rangle_{\sigma}=r^2\sin^2\theta ~V(r)\ ,\ \ 
    \langle \mathbf{g}_{11,11}\rangle_{\sigma}=V^{-1}(r)\ .\label{tnut}
\end{align}
We hasten to point out that the expectation value \eqref{delmet} is a correction to the background Minkowski metric \eqref{mink} and \eqref{tnut} is the fully corrected metric on ${\bf R}^{1,6}\times$ Taub-NUT space.  We only need to find the functions $\overline\alpha_{\rm MN}(k)$ for ${\rm M,N} = 7,8,9,11$ directions because  ${\bf R}^{1,6}$ metric agrees with the background Minkowski metric.  Utilising the spherical symmetry of the $7,8,9$ directions, the metric components in \eqref{tnut} are written in terms of $r$, $\theta$, and $\varphi$ coordinates with $x^{11}$ being a periodic direction with periodicity $4\pi\mu$.  

The functions $\overline\alpha_{\rm MN}(k)$ for ${\rm M,N}$ belonging to ${\bf R}^{1,6}$ simply vanish.  Our problem therefore has reduced to determining $\overline\alpha_{rr}(k)$, $\overline\alpha_{\theta\theta}(k)$, $\overline\alpha_{\varphi\varphi}(k)$, and $\overline\alpha_{11,11}(k)$ only.  The spherical symmetry of the Taub-NUT space in the $7,8,9$ direction implies we only need to determine one function $V(r)$ of the radial coordinate $r$.  The self-duality of the Riemann curvature guarantees that the component $\langle\mathbf{g}_{11,11}\rangle_\sigma = V^{-1}(r)$.  The components of $\overline\alpha_{\rm MN}(k)$ along 3D polar direction are
\begin{align}\label{alpsphere}
    \overline\alpha_{rr} &= \frac{\mu}{\pi^2}\delta(k_0)\delta^{6}(\tilde {\bf k})\delta(k_{11})\ ,\nonumber \\
    \overline\alpha_{\theta\theta} &= \delta(k_{0})\delta^{(6)}(\widetilde{\mathbf{k}})\delta(k_{11})\left(\frac{3}{16\pi} \hat{{k}}\delta^{'''}(\hat{{k}})- \frac{\mu}{\pi^2 {\hat{k}}^2} \right)\ ,    
\end{align}
and $\overline{\alpha}_{\varphi\varphi}$ may be easily determined from above; where
$k \equiv (k_0, \tilde{\bf k}, \hat{\bf k}, k_{11})$ with $(k_0, \tilde{\bf k})$ parametrizing the seven dimensional momenta along ${\bf R}^{1, 6}$; $(\hat{\bf k}, k_{11})$ parametrizing the momenta along the Taub-NUT base and eleventh direction respectively; and $\hat{\bf k} = (\hat{k} \sin\tilde\theta \cos \tilde\varphi, \hat{k} \sin \tilde\theta \sin \tilde\varphi, \hat{k} \cos\tilde\theta)$.
The component of metric in the eleventh direction is a circle fibration over the base with a function $V^{-1}(r)$.  The GS function for which is therefore more complicated and can be written in the following integral representation,
\begin{widetext}
    \begin{equation}\label{11circle}
     \overline{\alpha}_{11,11}(k)= \delta(k_{0})\delta^{(6)}(\Tilde{\bf k})\delta(k_{11})\left[\hat{\mathbf{k}}^2 \delta^{(3)}(\hat{\bf k})+ \frac{{\rm R}_{11}}{4 \pi^2 } \left(1  -\frac{\hat{k} {\rm R}_{11}}{2}\int_{{\rm R}_{11} \hat{k}}^{\infty}\frac{\sin~t}{t- \hat{k} {\rm R}_{11}/2} ~dt  \right)-\frac{1}{4\pi}\hat{k}\delta^{\prime\prime \prime}(\hat{k})\right].
\end{equation}
\end{widetext}
Although we have shown that the eleven dimensional Fourier integral in \eqref{resurmet} completely reproduces the M-theory KK monopole solution, we have not computed the contribution of the principal value integral $\textstyle\mathbb{I}_{c,l}$ as yet.  Note that this integral is an outcome of summing over all perturbative as well as non-perturbative contributions to the effective action. One might worry that, in the absence of the precise knowledge of the nodal diagram amplitude ${\cal A}$, the principal value of the integral will be harder to compute because of the  dependence of the pole on ${\cal A}$ in the Borel plane. Fortunately, this is not the case because the coupling $g$ that appears in the definition of 
$c \equiv \left({\cal A} g\right)^{1/l}$ in \eqref{delmet} takes the form
\vspace{-2mm}
\begin{equation}
    g \propto \frac{1}{{\rm M}_{p}^{+ve}}\ ,
\end{equation}
where the positive power is determined by the derivative coupling in the action. In the low energy limit, {\it i.e.} when ${\rm M}_{p}\to\infty$,  we are in the weak coupling limit and hence $g\to 0$ which implies that $c\to 0$.  In this limit the principal value integral in \eqref{delmet} becomes \cite{Brahma4}:
\begin{equation} \label{limiting} \lim_{c \to 0} \frac{\mathbb{I}_{c, l}}{c} ~ \to ~ 1, 
\end{equation}
implying that the entire resurgent sum, including the non-perturbative terms, has only a constant contribution with no sub-leading polynomial or exponentially suppressed terms.  Thus the principal value integral {\it does not affect} the result obtained by doing the eleven dimensional Fourier transform. Adding the NLO diagrams, or other higher derivative interactions in ${\bf S}_{\rm tot}$, do not change the aforementioned conclusion \cite{Brahma5}.

Before we show that this solution is a seed to incorporate all string theory solitons as GS coherent states, we will establish another result that will be useful in the case of string soliton solutions.  The Minkowski vacuum solution preserves all the supersymmetry of the M-theory because it is a solution with trivial holonomy group.  The M-theory KK monopole solution is the next simplest solution.  The Taub-NUT space is a non-compact hyperk\"ahler four manifold.  That is it is Ricci flat and solves vacuum Einstein equations. (In the language of expectation values, the GS state satisfies the Schwinger-Dyson's equations \cite{Brahma1, Brahma2, Brahma3}.) The self-duality condition implies the spin connection is half flat and hence the  space has a $SU(2)$ holonomy.  The 32 component Majorana spinor in eleven dimensions $-$ by combining $\Psi_{{\rm M}\alpha}$ and $\overline\Psi_{{\rm M}\alpha}$ $-$ decomposes as $32=8\times 4$, with 8 component spinor along ${\bf R}^{1,6}$ direction and 4 component spinor along the Taub-NUT direction.  The trivial holonomy along ${\bf R}^{1,6}$ implies all 8 components of the spinor are preserved but $SU(2)$ holonomy of the Taub-NUT space implies only 2 out of 4 components are preserved.  We therefore get 16 Killing spinors characterising this space and hence the GS state corresponding to the M-theory KK monopole is a half-BPS state.

\mypara{\textbf{Solitons in String Theory:}} \label{solstr} 
In the above section, we showed how to realise the M-theory KK monopole as a GS coherent state.  We will now show that all the string theory soliton solutions can be derived from this solution by using a sequence of dimensional reductions and string theoretic duality transformations.  For example, by choosing any of the space-like directions in the ${\bf R}^{1,6}$ direction to be the M-theory circle, one obtains the KK monopole solution in type IIA theory.  We could have arrived at this solution without taking the M-theory route.  That is, we could have directly solved for the GS functions in type II theory (A as well as B) to get identical results.  The type IIB KK monopole is just one T-duality, along a space-like direction in ${\bf R}^{1,5}$, away from type IIA KK monopole.  On the other hand, if we T-dualise the Taub-NUT circle direction of type II(A/B) KK monopole, we end up with the NS5-brane of type II(B/A) theory respectively.

If instead we chose the Taub-NUT circle direction to be the M-theory direction then we end up with the D6-brane solution of type IIA string theory.  It is then trivial to generate all Dp-brane solutions by T-duality transformations.  Since the initial seed is the KK monopole solution to the M-theory equations of motion, its coherent state representation along with duality cascade guarantees that all string theory KK monopoles, NS5-branes as well as any Dp-brane can also be written in terms of the GS coherent state.  All we need to do is carefully follow the chain of duality transformations.  Recall that the GS function along the worldvolume of the KK monopole vanishes and it is nontrivial along the Taub-NUT direction.  Therefore from the GS state perspective, T-duality transformation along the D-brane worldvolume corresponds to simply switching from one coherent state shift function to another, {\it i.e.} $\mathbb{D}(\sigma_1)\vert\Omega\rangle \to \mathbb{D}(\sigma_2)\vert\Omega\rangle$, and T-duality in the transverse direction will be the reverse process, namely $\mathbb{D}(\sigma_2)\vert\Omega\rangle \to \mathbb{D}(\sigma_1)\vert\Omega\rangle$.

Since dimension reduction as well as the T-duality transformations along the isometry direction do not affect the number of Killing spinors, all these solutions are also half-BPS solutions in string theory and hence the GS coherent state is a half BPS state.  

The half BPS solutions possess no force property which allows us to stack them one over the other or keep them separated without affecting the Killing spinors.  The GS ansatz for the coherent state naturally incorporates this fact to produce multi-Taub-NUT solution.  All we need to do is to replace, for example:
\begin{equation}\label{replace}
\overline\alpha_{rr} ~~ \to ~~ \overline\alpha_{rr}\times \Big(1 + \sum\limits_{j=1}^{n-1}\exp(ikr_j)\Big), 
\end{equation}
and we get multi-Taub-NUT solution with one solution centred at $r=0$ and remaining $n-1$ of them centred at $r=r_j$ one at each $j$.  This simple additive ansatz for multi-soliton seems to suggest that the GS function $\overline\alpha_{\rm MN}$ may be obtained by using some solution generating technique. We hope to address this issue in a forthcoming work.

Before closing this section, let us look at yet another pure gravity solution to M-theory or string theory equations of motion.  In the M-theory this solution corresponds to the geometry ${\bf R}^{1,2}\times \mathrm{TN}\times \mathrm{TN}$, where $\mathrm{TN}\times \mathrm{TN}$ is two copies of Taub-NUT space extended along $x^3,\cdots x^9, x^{11}$ directions.  Recall in the ${\bf R}^{1,6}\times \mathrm{TN}$ the GS function $\overline\alpha_{\rm MN}$ along the ${\bf R}^{1,6}$ direction was trivial, namely zero.  All we need to do is to replace it with another copy of $\overline\alpha_{\rm TN}$ along $x^3,x^4,x^5,x^6$ directions.  This can be achieved with ease implying that ${\bf R}^{1,2}\times \mathrm{TN}\times \mathrm{TN}$ is a coherent state in M-theory and ${\bf R}^{1,1}\times \mathrm{TN}\times \mathrm{TN}$ is a coherent state in type II string theory.

To show that this is a quarter BPS solution, let us decompose 32 component Majorana spinor into $32= 2\times 4\times 4$, with 2 component spinor along ${\bf R}^{1,2}$ and one 4 component spinor each along two Taub-NUT spaces.  Since both Taub-NUT spaces preserve 2 component spinors, we have $2\times 2\times 2=8$ Killing spinors, implying that this is a quarter BPS solution and the corresponding GS state is a quarter BPS state.  The duality symmetries of string theory relate this solution to intersecting brane configurations and hence the quarter BPS GS state gives a coherent state representation of intersecting brane configurations.

\mypara{\textbf{Where are the Open Strings?:}} 
So far we have shown how to obtain the ground state configuration of these solitons in terms of the GS coherent state.  A natural question would be how to obtain excitations over these solutions.  In particular, how do we see that the D-branes as coherent states naturally realise the open string excitations.

We will restrict our demonstration to the D6 brane, although using duality symmetries identical analysis can be carried out for other branes as well.  Whereas an open fundamental string stretched between two different D-branes creates oppositely charged excitations on two different branes, an open string excitation on a single D-brane generates an electric dipole excitation on it.  In the case of multi-D6 brane configuration, Sen \cite{Sen:1997js, Sen:1997kz} has shown that the stretched string between two D6-branes becomes a wrapped 2-brane on a transverse two cycle from the M-theory perspective.  A natural question would be what is the representation of the fundamental string excitation on the D6-brane in the M-theory?  Since this string generates a dipole on the D6-brane, it must correspond to a dielectric membrane in the M-theory.  Such an excitation is generated by the Myers effect \cite{Myers}, which requires a four form flux as well as multiple D0-brane excitations.  The D0-branes are simply the KK momenta along the M-theory circle, which in the D6-brane context correspond to graviton modes along the Taub-NUT circle direction. 

The above discussion clearly has its root in the M(atrix) theory formulation of M-theory \cite{BFSS} as the open strings between parallel D6-branes would be generated from Myers effect \cite{Myers}. Again a detailed study of this is clearly beyond the scope of this work, but we want to end this section by comparing our analysis with the one done previously by Shenker \cite{Shenker}. In \cite{Shenker}, D-branes were identified from Borel resumming a Gevrey series coming from the computation of the correlation function of two operators in string perturbation theory. The asymptotic nature of the correlation function computation, done over the interacting vacua, suggested the presence of non-perturbative effects that go as ${\rm exp}\left(-\frac{n}{ g_s}\right)$ and  ${\rm exp}\left(-\frac{n}{ g^2_s}\right)$, amongst other sub-dominant renormalon contributions. The former is related to the D-branes and the latter to the NS-branes, and the open-string nature of the D-branes was shown later in \cite{Polchinski:1995mt}. 

The analysis presented here differs from \cite{Shenker} in many ways. First, our analysis is done over an excited state, {\it i.e} over a Glauber-Sudarshan state, and not over a vacuum state. Secondly, we are computing one-point functions of the graviton, flux and fermionic operators over the GS states. These one-point functions would vanish in the analysis of \cite{Shenker}. Thirdly, the open-string behavior of the D-branes in the M-theory picture appears exclusively from the KK modes of the gravitons in the presence of four-form G-fluxes $\langle {\bf G}_{\rm MNPQ}\rangle_\sigma$ due to Myers effect. This is clearly different from how the open strings appear in say \cite{Polchinski:1995mt}. In fact, the gauge theory on D6-brane, including the tension of the open-string, may be easily determined from the presence of the normalizable self-dual harmonic two-form in the Taub-NUT space following \cite{Imamura, Sen:1997js, Sen:1997kz}.
Finally, in \cite{Shenker} the resurgent sum was crucial in deciphering the presence of D-branes, while in our case the resurgent sum in say \eqref{delmet} only contributes as wave-function renormalization (which in the limit ${\rm M}_p \to \infty$ is simply a constant).

The above comparison however raises the following question. In Borel resumming the Gevrey series in our case, do we expect the M2 and M5 instantons to also appear? The answer is in principle yes, but if we view all non-perturbative states in M-theory as coming from puffed-up D0 branes in the presence of appropriate G-flux components, then taking the metric components (and their KK modes) we are in principle taking into account all non-perturbative states as well as the renormalons in the Borel resummation. Happily M(atrix) theory guarantees this, but it will be interesting to explore this further.

\mypara{\textbf{Gauge-Gravity Dualities and RG Flows:}} 
Our study of coherent states in M-theory now leads to a rather interesting application in the field of non-AdS/non-CFT dualities. The non-CFT, as the name suggests, corresponds to theories that have non-trivial RG flows, and here we will specifically look at the low energy dynamics of a certain large ${\rm N}$ gauge theory that has permanent confinement in the far IR. An example of such a theory is a pure glue $SU(N)$ gauge theory that comes from ${\rm N}$ D5-branes wrapping a two-cycle of the resolved conifold in type IIB theory. Such a configuration is mirror-dual to ${\rm N}$ D6-branes wrapping a three-cycle of a deformed conifold in type IIA theory \cite{vafa1, Maldacena:2000yy, Vafa2, Cachazo}. Lifting this to M-theory now leads to a seven-dimensional manifold with ${\rm G}_2$ holonomy. In fact, this manifold appears from non-trivially combining a multi Taub-NUT geometry with ${\rm N}$ coinciding centers, coming from the ${\rm N}$ wrapped D6-branes, and the deformed conifold. From our earlier analysis it is now easy to construct the GS state associated with this geometry as the metric structure of such a manifold has been worked out in \cite{Maldacena:2000yy, Atiyah, Becker:2004qh, Alexander:2004eq, Becker:2005ef}. Let us label this GS states as $\vert \sigma_{\rm GT}\rangle$, such that the metric may be computed as before as an expectation value $\langle {\bf g}_{\rm MN}\rangle_{\sigma_{\rm GT}}$. Interestingly, under a flop transition and a subsequent dimensional reduction, we get a resolved conifold with two-form fluxes threading the two-cycle of the resolved conifold \cite{Atiyah}. This then corresponds to the gravity dual of the wrapped D6-brane configuration \cite{vafa1, Maldacena:2000yy, Vafa2, Cachazo}. 

Something interesting happens now. In the dual side in M-theory, generated from the flop transition \cite{Atiyah}, the metric configuration is again known \cite{Becker:2004qh, Alexander:2004eq, Becker:2005ef}. We can use our aforementioned techniques to determine the GS state associated with the dual configuration. Let us label it now as $\vert \sigma'_{\rm GT}\rangle$ and express the metric for the dual configuration as $\langle {\bf g}_{\rm MN}\rangle_{\sigma'_{\rm GT}}$. Comparing the two GS states, it appears that the following deformation:
\begin{equation}\label{gt}
\vert\sigma_{\rm GT}\rangle ~~ \longrightarrow~~\vert\sigma'_{\rm GT}\rangle, \end{equation}
from one GS state to another would be our way to represent the gauge-gravity duality here! Saying it in another way, gauge-gravity duality may simply be represented by deforming the GS state. But how unique is this deformation? Clearly, supersymmetry plays a crucial role here and, since we are dealing with minimal ${\cal N} = 1$ supersymmetry in four-dimensions, any arbitrary deformation of the GS state  $\vert\sigma_{\rm GT}\rangle$ would generically break supersymmetry. On the other hand, if there exist multiple supersymmetry preserving deformations of the GS state, that would either imply multiple possible gravity duals (modulo diffeomorphisms or even set of T-dualities in IIA), or multiple possible configurations that would be connected by some transformations to the gravity dual. Unfortunately, both these possibilities have issues: the former can be ruled out from the uniqueness of the gravity dual and the latter can be ruled out from the fact that a pure glue theory has neither a Coulomb branch nor a Higgs branch. Thus it appears that \eqref{gt} could  be a unique deformation (modulo diffeomorphisms or even number of T-dualities in the IIA side). More details will be presented elsewhere.

\mypara{\textbf{Discussion:}} \label{sec:discuss} 
We have presented the KK monopole solution of the M-theory equations of motion in terms of a GS coherent state.  This construction works for the KK monopoles in all string theories (type I, type II, as well as heterotic).  It is shown that using duality symmetry in string theory, a GS state representation for all solitons (D-branes and NS-branes) in string theory can be obtained.  Implementation of T-duality in the GS state simply corresponds to a specific swap between one GS state to another. Interestingly the gauge/gravity duality for pure glue theories may also be explained by yet another set of susy preserving deformations of the GS states.  Although our approach gives the GS representation of all solitonic solutions in string and M-theory, it would be interesting to relate this to a solution generating method that allows us to get multi-soliton solutions.

\noindent{\bf Acknowledgments:} DJ, DPJ, AM, and RT would like to thank Physics department, McGill University for warm hospitality. The work of JC is supported by the Simons Collaboration on Nonperturbative Bootstrap.  The work of KD is supported in part by a Discovery Grant from the Natural Sciences and
Engineering Research Council of Canada (NSERC). The work of AM is supported in part by the Prime Minister’s Research Fellowship provided by the Ministry of Education, Government of India.

\bibliographystyle{apsrev4-1}
\bibliography{references.bib}
\end{document}